\documentclass[prd,%
preprint,%
tightenlines,showpacs,%
nofootinbib,%
amsfonts,amsmath]{revtex4}

\begin{document}

\title{Arbitrary $p$-form Galileons}

\author{C.~Deffayet} \email{deffayet@iap.fr}
\affiliation{AstroParticule \& Cosmologie,
UMR 7164-CNRS, Universit\'e Denis Diderot-Paris 7,
CEA, Observatoire de Paris,
10 rue Alice Domon et L\'eonie
Duquet, F-75205 Paris Cedex 13, France,\\
and Institut d'Astrophysique de Paris, UMR 7095-CNRS,
98bis boulevard Arago, F-75014 Paris, France}

\author{S.~Deser} \email{deser@brandeis.edu}
\affiliation{Physics Department, Brandeis University, Waltham
MA 02454, USA, and Lauritsen Laboratory, California
Institute of Technology, Pasadena CA 91125, USA}

\author{G.~\surname{Esposito-Far\`ese}} \email{gef@iap.fr}
\affiliation{${\mathcal{G}}{\mathbb{R}}
\varepsilon{\mathbb{C}}{\mathcal{O}}$, Institut d'Astrophysique
de Paris, UMR 7095-CNRS, Universit\'e Pierre et Marie
Curie-Paris 6, 98bis boulevard Arago, F-75014 Paris, France}

\begin{abstract}
We show that scalar, $0$-form, Galileon actions ---models whose
field equations contain only second derivatives--- can be
generalized to arbitrary even $p$-forms. More generally, they
need not even depend on a single form, but may involve mixed $p$
combinations, including equal $p$ multiplets, where odd
$p$-fields are also permitted: We construct, for given dimension
$D$, general actions depending on scalars, vectors and higher
$p$-form field strengths, whose field equations are of exactly
second derivative order. We also discuss and illustrate their
curved-space generalizations, especially the delicate non-minimal
couplings required to maintain this order. Concrete examples of
pure and mixed actions, field equations and their curved space
extensions are presented.
\end{abstract}

\date{July 29, 2010}
\preprint{CALT 68-2792, BRX TH-620}

\maketitle

\section{Introduction}
\label{sec1}
The geometric ancestors of Galileons~\cite{Fairlie,%
Nicolis:2008in,DEFV09,Deffayet:2009mn,deRham:2010eu}
are the Gauss-Bonnet-Lovelock (GBL) actions
\begin{equation}
I=\int d^D x\, \varepsilon^{\mu\nu\dots}
\varepsilon^{\alpha\beta\dots}\,
R_{\mu\nu\alpha\beta} \dots R_{\dots}\,
e_{\rho\gamma} \dots e_{\dots},
\label{eq1}
\end{equation}
powers of the curvature $R$ whose field equations are
nevertheless independent of higher than second metric
derivatives. This is achieved by virtue of the Bianchi
identities, due to which the $R$-variations do not contribute;
only the explicit vielbeins' do, as is especially clear in
vielbein/spin connection formalism. Here $R$ is the ``field
strength'' of the (non-Abelian) spin connection
$\omega_{\mu\alpha\beta}(e)$; the Levi-Civita symbol
$\varepsilon^{\mu\nu\dots}$ is a tensor density, while
$\varepsilon^{\alpha\beta\dots}$ is a world scalar; $(\mu, \nu,
\ldots)$ and $(\alpha, \beta, \ldots)$ are world and local
Lorentz indices respectively. These actions are
dimension-dependent, yielding vanishing field equations below a
certain $D$, such as $D=5$ for $R^2$ and $D=7$ for $R^3$. More
explicitly, for $D=5$ say, one $e_{\mu\alpha}$ is required to
contract the two leftover indices in $(\varepsilon \varepsilon R
R)^{\mu\alpha}$, while there is no $e_{\mu\alpha}$, hence no
field equation, in $D=4$. The mechanism is simple, and as we
shall see below, universal: First note that $\delta R_
{\mu\nu\alpha\beta} = {\cal D}_{[\mu} \delta \omega_
{\nu]\alpha\beta}$, where ${\cal D}$ is the usual covariant
derivative with respect to the spin connection (acting also on
local indices), and $\delta \omega$ is a world vector. Therefore,
integrating ${\cal D}$ by parts (freely past all vielbeins of
course) onto the remaining Riemann tensor(s) gives $0$ by the
cyclic Bianchi identities. So the GBL field equations,
$(\varepsilon \varepsilon R\ldots R)^{\mu\alpha}=0$, just result
from removing (any) $e_{\mu\alpha}$ in (\ref{eq1}) and are
manifestly independent of higher than second vielbein
derivatives.

Galileons are scalars whose field equations depend only on second
derivatives, hence are invariant under constant shifts of the
fields (``positions'') and their gradients (``velocities''),
recalling old-fashioned Galilean invariance. [Note that this
invariance is only meaningful in flat space, since there are no
constant vectors or tensors in curved backgrounds.] Their actions
bear a formal resemblance to the GBL systems, when expressed
as~\cite{Deffayet:2009mn}
\begin{equation}
I=\int d^D x\, \varepsilon^{\mu\nu\dots}
\varepsilon^{\alpha\beta\dots}\,
\partial_\mu\pi \partial_\alpha\pi\,
(\partial_\nu\partial_\beta\pi)\dots(\partial\partial\pi).
\label{eq2}
\end{equation}
Again, the variations only leave second order equations
$(\partial\partial\pi)\dots(\partial\partial\pi)=0$, and for
sufficiently low $D$, where the $\partial\pi \partial\pi\sim
``e"$ are absent (for a given power of
$\partial\partial\pi\sim``R"$), Eq.~(\ref{eq2}) has vanishing
variation. This (slightly imperfect) similarity led us to
conjecture that (\ref{eq2}) could be obtained from a GBL-like
action using a metric suitably parametrized in terms of
$\partial\pi$; it was indeed elegantly confirmed
recently~\cite{deRham:2010eu} for the $R^2$ case (and in a
suitable limit), as a byproduct of a brane analysis.

The purpose of this Letter is to generalize the above models by
noting that the properties of $0$-forms underlying (\ref{eq2})
are actually shared by arbitrary even $p$-forms, and extend to
any (dimensionally allowed) admixtures of various $p$-level
fields. Surprisingly, we found a fundamental divide between even
(scalars, \dots) and odd (vectors, \dots) models. The latter turn
out, despite initial appearances, to have empty flat space
actions\footnote{Nevertheless, covariantized versions of trivial
flat space actions can produce nontrivial field equations,
proportional to the curvature, as will be illustrated in
Sec.~\ref{sec4}.} (except of course the standard Maxwell-like
${\cal L}=F^2$), i.e., devoid of field equations for any ($D$,
$p=2n+1$). However, as we discuss below, they may appear in mixed
form, or in multiplets of single $p$-form, models.

We will work primarily in flat space in order to focus on our
main results. As for scalars, the key ingredient here is that the
forms'``field strengths'' $\omega_{p+1}= dA_p$, are curls which
do not become covariant; only the explicit $\nabla$ in
$\nabla\omega$ does. Using these ``gauge-invariant'' field
strengths rather than ordinary gradients is both essential to the
Galileon aspect and excludes their, possible ghost, lower spin
gauge components.

Retaining second order upon extension to curved backgrounds is
nontrivial; even for scalars, the minimal coupling extension of
(\ref{eq2}) gave rise to third derivative terms in its stress
tensor and hence in the associated gravitational field equations,
as well as to third metric derivative terms $\propto \nabla R$ in
the $\pi$ equations. A delicate set of additional, non-minimal,
couplings, involving the full curvature tensor was
required~\cite{DEFV09,Deffayet:2009mn} to remove these. This
program, though correspondingly more complicated, can in fact be
carried out for our present generalized framework in a fashion
similar to that of \cite{Deffayet:2009mn} for scalars. Instead of
detailing here the straightforward but still rather lengthy
derivation for the most general case, we will display some
examples of successful covariantization in Sec.~\ref{sec4}.

\section{$p$-form actions}
\label{sec2}
We start, to emphasize the pitfalls in this problem, with the
obviously simplest ---but actually empty--- generalization from a
scalar (\ref{eq2}) to a one-form $A_\mu$ with field strength
$F_{\mu\nu}=\partial_{[\mu}A_{\nu]}$:
\begin{equation}
I= \int d^D x\, \varepsilon^{\mu\nu\dots}
\varepsilon^{\alpha\beta\dots}\,
F_{\mu\nu} F_{\alpha\beta}
(\partial_\rho F_{\gamma\delta}\, \dots)
(\partial_\epsilon F_{\sigma\tau}\, \dots),
\label{eq3}
\end{equation}
where the parentheses contain products of $\partial F$ and
indices are connected as follows: In the first parenthesis, the
index of the derivative $\partial$ is contracted with the first
$\varepsilon^{\mu\nu\dots}$ whereas those of $F$ are contracted
with the second $\varepsilon^{\alpha\beta\dots}$, and inversely
in the second parenthesis. The integrand of (\ref{eq3}) is a
total divergence which we may write as
$\frac{1}{2}\partial_\epsilon
[\varepsilon^{\mu\nu\dots\varphi\chi\dots}
\varepsilon^{\alpha\beta\dots\epsilon\zeta\dots}\, F_{\mu\nu}
F_{\sigma\tau} F_{\alpha\beta} (\partial_\rho F_{\gamma\delta}\,
\dots) (\partial_\zeta F_{\varphi\chi}\, \dots)]$. This equality
follows by noting that $\frac{1}{2}\partial_\epsilon$ manifestly
annihilates all its operands but $F_{\mu\nu} F_{\sigma\tau}$ (on
each of which it acts identically), where its actions reproduce
the original Lagrangian. Since this conclusion is due to the
evenness of $F_{\mu\nu} F_{\sigma\tau}$ upon exchange of their
indices $\mu\nu \leftrightarrow \sigma\tau$, the difficulty
obviously persists for all ($D$, $p=2n+1$). However, as we shall
see in the next section, odd $p$ models can be revived if they
are allowed to depend on more than one $A_\mu$.

Fortunately, the direct extension of (\ref{eq2}) does exist for
even $p$; it formally resembles (\ref{eq3}) where now
$\omega_{\lambda\mu\nu\dots} = \partial_{[\lambda}
A_{\mu\nu\dots]}$. In detail, the general (now non-vanishing)
action is
\begin{equation}
I=\int d^D x\, \varepsilon^{\mu\nu\dots}
\varepsilon^{\alpha\beta\dots}\,
\omega_{\mu\nu\dots} \omega_{\alpha\beta\dots}
(\partial_\rho\omega_{\gamma\delta\dots}\, \dots)
(\partial_\epsilon \omega_{\sigma\tau\dots}\, \dots).
\label{eq4}
\end{equation}
As in Eq.~(\ref{eq3}), when the derivative $\partial$ of a
gradient $\partial\omega$ is contracted with one of the two
$\varepsilon$, the indices of $\omega_{p+1}$ must be contracted
with the other $\varepsilon$, otherwise the action would vanish
by virtue of the Bianchi identities (i.e., $[d,d]=0$). The two
parentheses of (\ref{eq4}) must contain the same number of terms,
not greater than $(D-p-1)/(p+2)$. [In the lowest, $p=0$, case, an
odd total number of $\partial\omega$ is also permitted,
cf.~(\ref{eq2}).] For fewer terms, there remain extra open
indices on each $\varepsilon$ that must be contracted between
them (using vielbeins in curved space, of course); this yields
(up to an overall factor) the same Lagrangian as in the lowest
possible dimension. There is actually a formal resemblance
between (\ref{eq4}) and (\ref{eq2}) that emerges from considering
the $(p+1)$ indices of the form $\omega$ as a multi-index $M$
replacing the single index of $\pi_{,\mu}$. Then, just as each of
the two indices of $\partial_\alpha\partial_\mu\pi$ must be
contracted with different $\varepsilon$, here $\alpha$ and $M$ of
$\partial_\alpha\omega_M$ must belong to different $\varepsilon$.

It should be clear from our notation and the Bianchi identities
that the field equations depend homogeneously on $\partial
\omega$ and not at all on $\omega$, hence they enjoy the
corresponding Galilean invariance, this time under a shift of
($A_{\mu\nu\dots}$, $\omega_{\mu\nu\rho\dots}$) by constants
antisymmetric tensors ($c_{[\mu\nu\dots]}$,
$k_{[\mu\nu\rho\dots]}$). For completeness, let us run through
the argument, entirely akin to those for gravity and scalars:
First, if either pure $\omega$ is varied, the explicit curl on
its $A$ can only land on the other $\omega$, since Bianchi
annihilates any $\partial_\mu \partial_{[\alpha}
\omega_{\beta\gamma\dots]}$ (or $\partial_\alpha \partial_{[\mu}
\omega_{\nu\rho\dots]}$). Likewise, varying any $\partial
\omega\sim \partial_\mu\partial_\alpha A$ factor forces each of
those two $\partial$ to land on one of the two pure $\omega$; all
other landings vanish, again by Bianchi. The simplest version of
Eq.~(\ref{eq4}), valid for $D\geq p+1$, does not contain any
derivative $\partial \omega$, and is thus the standard kinetic
term $\omega^2$ (valid for all $p$ of course, though dynamically
non-trivial only when $D> p+1$). The first novel $p$-form
Galileon action, involving just two $\partial \omega$ factors,
requires $D \geq 2p+3$. For $p = 2$, it reads explicitly
\begin{eqnarray}
I&=&\int d^7 x\, \varepsilon^{\mu\nu\rho\sigma\tau\varphi\chi}
\varepsilon^{\alpha\beta\gamma\delta\epsilon\zeta\eta}\,
\omega_{\mu\nu\rho}\, \omega_{\alpha\beta\gamma}\,
\partial_\sigma \omega_{\delta\epsilon\zeta}\,
\partial_\eta \omega_{\tau\varphi\chi}
\nonumber\\
&=& 36\int d^7 x \Bigl[
- 9 (\omega^\mu_{\hphantom{\mu}\nu\rho,\sigma}
\omega^{\sigma\tau\varphi}
\omega_{\tau\varphi\mu,\chi} \omega^{\chi\nu\rho})
- 18 (\omega^{\mu\hphantom{\nu}\rho}_{\hphantom{\mu}\nu}
\omega_{\mu\sigma}^{\hphantom{\mu\sigma}\tau}
\omega^{\varphi\chi\nu,\sigma} \omega_{\varphi\chi\tau,\rho})
\nonumber\\
&&- 36(\omega^{\mu\nu\rho} \omega_{\rho\sigma\tau}
\omega_{\mu\nu\varphi}^{\hphantom{\mu\nu\varphi},\sigma}
\omega^{\tau\varphi\chi}_{\hphantom{\tau\varphi\chi},\chi})
+ 6 (\omega_{\mu\nu\rho} \omega^{\mu\nu\rho,\sigma}
\omega_{\sigma\varphi\chi}
\omega^{\varphi\chi\tau}_{\hphantom{\varphi\chi\tau},\tau})
+ 18 (\omega_{\mu\nu}^{\hphantom{\mu\nu}\rho}
\omega^{\mu\nu\sigma} \omega_{\varphi\chi\rho,\sigma}
\omega^{\varphi\chi\tau}_{\hphantom{\varphi\chi\tau},\tau})
\nonumber\\
&&- 3 (\omega^{\mu\nu\lambda} \omega_{\rho\sigma\tau,\lambda})^2
- 9 (\omega^{\mu\nu\rho} \omega_{\rho\sigma\tau,\lambda})^2
+ 18 (\omega_{\mu\nu\rho}
\omega^{\rho\sigma\tau}_{\hphantom{\rho\sigma\tau},\tau})^2
+ 9 (\omega^{\mu\nu\rho} \omega_{\mu\nu\sigma,\tau})^2
\nonumber\\
&&- 9 (\omega_{\mu\nu\rho}
\omega^{\mu\nu\sigma}_{\hphantom{\mu\nu\sigma},\sigma})^2
- (\omega^{\mu\nu\rho} \omega_{\mu\nu\rho,\sigma})^2
+ (\omega^2) (\omega_{\mu\nu\rho,\sigma})^2
- 3 (\omega^2)
(\omega^{\mu\nu\rho}_{\hphantom{\mu\nu\rho},\rho})^2
\Bigr].
\label{eq5}
\end{eqnarray}
Its field equation
\begin{equation}
\varepsilon^{\mu\nu\rho\sigma\tau\varphi\chi}
\varepsilon^{\alpha\beta\gamma\delta\epsilon\zeta\eta}\,
\partial_\rho \omega_{\alpha\beta\gamma}\,
\partial_\sigma \omega_{\delta\epsilon\zeta}\,
\partial_\eta \omega_{\tau\varphi\chi}= 0
\label{eq6}
\end{equation}
is obviously of pure second order; we do not display its, 23
term, expansion.

We conclude this section by discussing another amusing (if
somewhat tangential) parallel between tensors and forms, which
evokes the well-known conversion \cite{MacDowell:1977jt} of pure
divergence $D=4$ GB into general relativity (GR), with or without
cosmological term: upon adding $\pm \Lambda e_{\mu\alpha}
e_{\nu\beta}$ to each $R_{\mu\nu\alpha\beta}$ in the topological
invariant (\ref{eq1}), it becomes proportional to the GR action,
since the cross-term in $\varepsilon \varepsilon (R \pm \Lambda e
e)^2$ is the scalar curvature, while the $\Lambda^2$ term is the
volume density, cosmological, term. Subtracting $(R+\Lambda
ee)^2$ and $(R-\Lambda ee)^2$ actions removes the latter. For
scalars, we similarly add $\pm m^2 \pi \eta_{\mu\alpha}$ to each
of the two $\partial_\mu\partial_\alpha\pi$ in the pure GB-like
$I=\int \varepsilon \varepsilon\, \partial\partial\pi\,
\partial\partial\pi$: this leads to the (massive or massless by
subtraction) Klein-Gordon action. These extensions can be made
for all forms: thus for the vector Proca/Maxwell actions, add
$\sim \pm m^2 \eta_{\mu[\alpha} A_{\beta]}$ to each $\partial_\mu
F_{\alpha\beta}$ in the otherwise vacuous action $I=\int
\varepsilon\varepsilon \partial F \partial F$, etc. Note that our
mass construction is valid in all $D$ [$\geq (p+2)$ of course];
that of GR, for all $D \geq 4$. That the former has a curved
space extension is also obvious.

\section{Mixed form actions}
\label{sec3}
Our actions (\ref{eq4}) can be further generalized by including
several species, i.e., mixtures of various unequal $p$-forms
compatible with a desired $D$. Labelling these species by
$(a, b, \dots)$, the action takes the formal expression
\begin{equation}
I=\int d^D x\, \varepsilon^{\mu\nu\dots}
\varepsilon^{\alpha\beta\dots}\,
\omega^a_{\mu\nu\dots} \omega^b_{\alpha\beta\dots}
(\partial_\rho\omega^c_{\gamma\delta\dots}\,\dots)
(\partial_\epsilon\omega^d_{\sigma\tau\dots}\,\dots).
\label{eq7}
\end{equation}
The number of indices contracted with the first and second
$\varepsilon$ must be the same and not greater than $D$, but the
two parentheses may now involve different species and therefore a
different number of terms. Here, Bianchi again ensures (exactly
as for the single species version) that only $\partial\omega$
appears in the field equations. Hence (flat space) Galilean
invariance under translation of all $(A_p, \omega_{p+1})$ by
constant antisymmetric tensors $(c_p, k_{p+1})$ is preserved.
Note that odd forms are also allowed in (\ref{eq7}), subject to
various symmetry constraints. For example, no more than two
$\partial F$ factors can be present, otherwise (\ref{eq7}) would
involve at least one product of the form $\partial_\mu
F_{\alpha\beta} \partial_\nu F_{\gamma\delta}$, where the two
$\partial$ and the two $F$ are respectively contracted with the
same $\varepsilon$ tensors. Hence their indices can be
interchanged by three permutations, $\mu\leftrightarrow \nu$,
$\alpha\beta\leftrightarrow \gamma\delta$, so they vanish
identically. This single $\partial F \partial F$ ceiling
obviously also applies to higher odd $p$-forms; instead, the
$p=2n$ models, being even under such permutations, may contain
arbitrary powers of $\partial\omega$ consistent with a given $D$
(but conversely, see below for limitations on even $p$ actions).

Let us quote two simple, mixed $0$ \& $1$-form, nontrivial
examples; the first Lagrangian is defined in any $D\geq 3$:
\begin{equation}
{\cal L} = \varepsilon^{\mu\nu\rho}
\varepsilon^{\alpha\beta\gamma}\,
F_{\mu\nu} F_{\alpha\beta}\,
\partial_\rho\partial_\gamma\pi
= 4 F^{\mu\rho} F^\nu_{\hphantom{\nu}\rho} \pi_{,\mu\nu}
-2 F^2\, \Box\pi.
\label{eq8}
\end{equation}
Both its $\pi$ and $A_\lambda$ field equations are obviously of
pure second order; explicitly,
\begin{eqnarray}
(F_{\mu\nu,\rho})^2
- 2 (F^{\mu\nu}_{\hphantom{\mu\nu},\nu})^2 &=& 0,
\label{eq9}\\
F^{\lambda\mu,\nu} \pi_{,\mu\nu}
+ F^{\mu\nu}_{\hphantom{\mu\nu},\nu}
\pi^{,\lambda}_{\hphantom{,\lambda}\mu}
- F^{\lambda\mu}_{\hphantom{\lambda\mu},\mu}\, \Box\pi
&=& 0.
\label{eq10}
\end{eqnarray}
Similarly, in $D \geq 4$, the mixed model
\begin{eqnarray}
{\cal L} &=& \varepsilon^{\mu\nu\rho\sigma}
\varepsilon^{\alpha\beta\gamma\delta}\,
\partial_\mu\pi \partial_\alpha\pi\,
\partial_\nu F_{\beta\gamma}\,
\partial_\delta F_{\rho\sigma}\nonumber\\
&=&- 8 (\pi_{,\mu} F^{\rho\mu,\nu}
F_{\rho\sigma}^{\hphantom{\rho\sigma},\sigma} \pi_{,\nu} )
+ 4 (\pi^{,\mu} F_{\mu\nu,\rho})^2
+ 2 (\pi^{,\mu} F_{\nu\rho,\mu})^2
\nonumber\\
&&-4 (\pi_{,\mu} F^{\mu\nu}_{\hphantom{\mu\nu},\nu})^2
- 2 (\pi_{,\mu})^2 (F_{\nu\rho,\sigma})^2
+ 4 (\pi_{,\mu})^2 (F^{\nu\rho}_{\hphantom{\nu\rho},\rho})^2
\label{eq11}
\end{eqnarray}
also yields pure second order $\pi$ and $A_\lambda$ field
equations:
\begin{eqnarray}
4 (\pi_{,\mu\nu} F^{\rho\mu,\nu}
F_{\rho\sigma}^{\hphantom{\mu\sigma},\sigma})
- 2 (F^{\mu\rho,\sigma} \pi_{,\mu\nu}
F^\nu_{\hphantom{\nu}\rho,\sigma})
+ 2 (F^{\mu\rho}_{\hphantom{\mu\rho},\rho} \pi_{,\mu\nu}
F^{\nu\sigma}_{\hphantom{\nu \sigma}, \sigma})&&
\nonumber\\
- (F_{\rho\sigma,\mu} \pi^{,\mu\nu}
F^{\rho\sigma}_{\hphantom{\rho\sigma},\nu})
+ (\Box\pi) (F_{\mu\nu,\rho})^2
- 2 (\Box\pi) (F^{\mu\nu}_{\hphantom{\mu\nu},\nu})^2 &=& 0,
\label{eq12}\\
2 (\pi_{,\mu\rho} F^{\lambda\mu}_{\hphantom{\lambda\mu},\nu}
\pi^{,\nu\rho})
+ 2 (\pi^{,\lambda\mu}F_{\mu\nu,\rho} \pi^{,\nu\rho})
+ 2 (\pi^{,\lambda\rho} \pi_{,\rho\mu}
F^{\mu\nu}_{\hphantom{\mu\nu},\nu})
- (\pi_{,\mu\nu})^2
(F^{\lambda\rho}_{\hphantom{\lambda\rho},\rho})&&
\nonumber\\
- 2 (\Box\pi) (\pi_{,\mu\nu} F^{\lambda\mu,\nu})
- 2 (\Box\pi) (\pi^{,\lambda}_{\hphantom{,\lambda}\mu}
F^{\mu\nu}_{\hphantom{\mu\nu},\nu})
+ (\Box\pi)^2 (F^{\lambda\mu}_{\hphantom{\lambda\mu},\mu})&=& 0.
\label{eq13}
\end{eqnarray}

An even simpler class of mixed actions involves a single
$p$-order species, but now as a ``multiplet''
$A_{\mu\nu\dots}^a$, for instance pure scalars but with different
$\pi^a$ replacing the single one in (\ref{eq2}). This extension
even resuscitates odd-$p$ actions: For instance, the simplest
bi-vector Lagrangian of the type (\ref{eq3}), ${\cal L}=
\varepsilon^{\mu\nu\rho\sigma\tau}
\varepsilon^{\alpha\beta\gamma\delta\epsilon}\, F^a_{\mu\nu}
F^a_{\alpha\beta}\, \partial_\rho F^b_{\gamma\delta}
\partial_\epsilon F^b_{\sigma\tau}$, is obviously no longer a
total divergence. Our reasoning below Eq.~(\ref{eq4}), showing
that the field equations do not involve higher order derivatives,
may also be generalized to non-Abelian gauge bosons $A^a_\mu$ and
their field strengths $F = dA + A\wedge A$, although both the
invariances under constant shifts, $A^a_{\mu} \rightarrow
A^a_{\mu} + c^a_{\mu}$ and $F^a_{\mu\nu} \rightarrow F^a_{\mu\nu}
+ k^a_{[\mu\nu]}$, would then be lost. Indeed, if ${\cal D}$
denotes the covariant derivative with respect to the internal
space (like the ${\cal D}$ below (\ref{eq1}) with respect to
tangent space), then the Bianchi identities ${\cal
D}^{\vphantom{a}}_{[\mu}F^a_{\nu\rho]} = 0$ still hold, therefore
Lagrangians of the form ${\cal L}= \varepsilon^{\mu\nu\dots}
\varepsilon^{\alpha\beta\dots}\, F^a_{\mu\nu} F^b_{\alpha\beta}
({\cal D}_\rho F^c_{\gamma\delta}\, \dots) ({\cal D}_\epsilon
F^d_{\sigma\tau}\, \dots)$ define non-linear extensions of
Yang-Mills theory, while keeping field equations of second (and
lower) order.

It is worth noting that one may also add undifferentiated powers
of $\omega$ beyond the two in the generalized models (\ref{eq7}),
provided all indices of any one $\omega$ (whatever its $p$-order)
are contracted with those of a single $\varepsilon$ tensor, but
not ``across'' both. Also, no more than two undifferentiated even
$p$-field strengths $\omega_{p+1}^a$ are allowed for the same
species $a$, otherwise the action would vanish by oddness, while
any number of odd $p$ field strengths may be present. The same
reasoning as above indeed shows that no higher derivative than
$\partial\omega$ is generated in the field equations, i.e., that
they depend at most on second derivatives of the $p$-forms $A$.
On the other hand, these field equations now involve some pure
$\omega$ in addition to the usual $\partial\omega$ factors,
because at most two derivatives are generated by varying the
$\partial\omega$ terms of the action, so that they can act on at
most two of the undifferentiated $\omega$. Therefore, this
generalization with more than two pure $\omega$ results in a loss
of the ``velocity'' invariance $\omega \rightarrow \omega + k$. A
simple $D \geq 4$ example of this type is
\begin{eqnarray}
{\cal L} &=& \varepsilon^{\mu\nu\rho\sigma}
\varepsilon^{\alpha\beta\gamma\delta}\,
\partial_\mu\pi \partial_\alpha\pi\,
F_{\nu\rho} F_{\beta\gamma}\,
\partial_\sigma\partial_\delta\pi
\nonumber\\
&=&4 (\pi_{,\mu} F^{\mu\nu} \pi_{,\nu\rho}
F^{\rho\sigma} \pi_{,\sigma})
+ 8 (\pi_{,\mu} F^{\mu\nu} F_{\nu\rho}
\pi^{,\rho\sigma} \pi_{,\sigma})
+ 2 (F^2) (\pi_{,\mu} \pi^{,\mu\nu} \pi_{,\nu})
\nonumber\\
&&+ 4 (\pi_{,\mu})^2 (F^\nu_{\hphantom{\nu}\sigma}
F^{\rho\sigma} \pi_{,\nu\rho})
-4 (\pi_{,\mu} F^{\mu\nu} F_{\nu\rho} \pi^{,\rho}) (\Box\pi)
- 2 (\pi_{,\mu})^2 (F^2) (\Box\pi).
\label{eq14}
\end{eqnarray}
As is clearest from the first expression, its variations involve
both first and second (but no higher) derivatives of $\pi$ and
$A_\mu$.

\section{Gravitational coupling}
\label{sec4}
As stated in the Introduction, second-order preserving extension
of even the scalar flat space actions to curved backgrounds was a
rather complicated process, one that becomes more combinatorially
involved for higher forms. We content ourselves here with giving
the explicit non-minimal extensions for four of our cases
(\ref{eq5}), (\ref{eq8}), (\ref{eq11}), (\ref{eq14}), that avoid
higher derivatives in both the matter and gravitational (that is,
through $T_{\mu\nu}$) field equations. These terms are
constructed as for scalars in \cite{Deffayet:2009mn}: All
possible pairs of gradients, $\partial\omega^a \partial\omega^b$,
must be replaced by suitable contractions of the undifferentiated
$\omega^a \omega^b$ with the Riemann tensor, and added to the
minimally covariantized flat-space action with suitable
coefficients; somewhat more involved counting shows that they
require factors $\propto (p_a+1)(p_b+1)$, where $p_{a,b}$ denote
the orders of the forms $A^{a,b}$. One other difference in the $p
> 0$ construction is that $\nabla_\mu\omega_{\alpha\beta\dots}$
are to be distinguished from $\nabla_\alpha\omega_{\mu\nu\dots}$,
essentially because of their different $\varepsilon$-index
contractions, a distinction irrelevant to the original scalar,
$\pi_{;\mu\alpha} = \pi_{;\alpha\mu}$, case. One common feature
is that flat-space Galilean invariance is also not restorable by
consistent covariantization (nor should it be expected, absent
constant vectors or tensors in curved space): the equations now
necessarily depend on both second and first derivatives of the
fields. For (\ref{eq5}), the added terms are:
\begin{eqnarray}
\Delta I&=&-\frac{9}{4}\int d^7 x\,
\varepsilon^{\mu\nu\rho\sigma\tau\varphi\chi}
\varepsilon^{\alpha\beta\gamma\delta\epsilon\zeta\eta}\,
\omega_{\mu\nu\rho}\, \omega_{\alpha\beta\gamma}\,
\omega_{\lambda\sigma\tau}\,
\omega^\lambda_{\hphantom{\lambda}\delta\epsilon}\,
R_{\varphi\chi\zeta\eta}
\nonumber\\
&=&54 \int d^7 x \sqrt{-g} \Bigl[
24 (\omega_{\mu\nu\lambda} \omega^{\lambda\tau\rho}
\omega_{\tau\varphi\chi} \omega^{\varphi\chi\sigma}
R^{\mu\nu}_{\hphantom{\mu\nu}\rho\sigma})
+ 12 (\omega_{\lambda\tau\mu} \omega^{\lambda\tau\nu}
\omega_{\varphi\chi\rho} \omega^{\varphi\chi\sigma}
R^{\mu\rho}_{\hphantom{\mu\rho}\nu\sigma})
\nonumber\\
&&+ 4 (\omega^2) (\omega_{\lambda\mu\nu}
\omega^{\lambda\rho\sigma}
R^{\mu\nu}_{\hphantom{\mu\nu}\rho\sigma})
+ 12(\omega_{\mu\nu\varphi} \omega^{\mu\nu\chi}
\omega^{\tau\varphi\rho}
\omega_{\tau\chi\sigma} R_\rho^\sigma)
+ 18 (\omega_{\mu\nu\tau} \omega^{\mu\nu\rho}
\omega^{\varphi\chi\tau} \omega_{\varphi\chi\sigma}
R_\rho^\sigma)
\nonumber\\
&&- 10 (\omega^2) (\omega^{\mu\nu\rho}
\omega_{\mu\nu\sigma} R_\rho^\sigma)
-3(\omega_{\mu\nu\rho} \omega^{\mu\nu\sigma})^2 R
+ (\omega^2)^2 R
\Bigr].
\label{eq15}
\end{eqnarray}
Similarly, the mixed $D\geq 4$ example (\ref{eq11}) acquires the
terms
\begin{eqnarray}
\Delta{\cal L}&=&\varepsilon^{\mu\nu\rho\sigma}
\varepsilon^{\alpha\beta\gamma\delta}\,
\partial_\mu\pi \partial_\alpha\pi\,
F_{\lambda\nu} F^\lambda_{\hphantom{\lambda}\beta}\,
R_{\rho\sigma\gamma\delta}
\nonumber\\
&=& 2 \sqrt{-g}\Bigl[
- 2 (\pi_{,\mu} \pi_{,\nu} F_{\rho\lambda}
F_{\sigma}^{\hphantom{\sigma}\lambda}
R^{\mu\rho\nu\sigma})
+ 4 (\pi_{,\mu}F^{\mu\nu} F_{\nu\rho}
R^{\rho\sigma} \pi_{,\sigma})
+ 2 (\pi_{,\mu})^2 (F_{\nu\sigma}
F_{\rho}^{\hphantom{\rho}\sigma} R^{\nu\rho})
\nonumber\\
&&+ 2 (F^2) (\pi_{,\mu} R^{\mu\nu} \pi_{,\nu})
+ (\pi_{,\mu} F^{\mu\rho})^2 R
- (\pi_{,\mu})^2 (F^2) R
\Bigr].
\label{eq16}
\end{eqnarray}
In contrast to the above models, the mixed $D\geq 3$ example
(\ref{eq8}) and the ``non-Galileon'' Lagrangian (\ref{eq14})
actually require \textit{no} additional terms to preserve second
order, since they only contain a single vulnerable ---because
second order--- $\partial\partial\pi$ factor. It is clear by
inspection of (\ref{eq8}) and (\ref{eq14}) that all (covariant)
third derivatives arising from variations here always have the
form of a commutator $[\nabla,\nabla]$ acting on a $\partial A$,
that is a ---harmless--- curvature times first derivatives of
fields.

Our final model illustrates the observation made in our footnote
that actions trivial in flat space can have non-trivial,
dynamical, curvature-dependent extensions: Consider the vector
models (\ref{eq3}), or more generally actions (\ref{eq4}) for any
odd $p$, which are vacuous in flat space. Their minimal
covariantizations are clearly both nonvanishing and of third
order. However, one may also add appropriate non-minimal terms
that both remove the offending higher derivatives and remain
non-trivial. Indeed, the simplest case is the lowest Galileon
$D=5$ vector action,
\begin{eqnarray}
I&=&\int d^5 x\, \varepsilon^{\mu\nu\rho\sigma\tau}
\varepsilon^{\alpha\beta\gamma\delta\epsilon}\,
F_{\mu\nu} F_{\alpha\beta}\, \nabla_\rho F_{\gamma\delta}\,
\nabla_\epsilon F_{\sigma\tau}
\nonumber\\
&=&-\frac{1}{2}\int d^5 x\, \varepsilon^{\mu\nu\rho\sigma\tau}
\varepsilon^{\alpha\beta\gamma\delta\epsilon}\,
F_{\mu\nu} F_{\alpha\beta}\,
F^\lambda_{\hphantom{\lambda}\rho} F_{\delta\gamma}\,
R_{\sigma\tau\lambda\epsilon}.
\label{eq17}
\end{eqnarray}
The last equality in (\ref{eq17}) exhibits the model's
curvature-dependence, and is obtained from the first expression
by parts integration. [The metric variation of the curvature in
the second expression (\ref{eq17}) yields a non-vanishing
$T^{\mu\nu} = \partial_\alpha \partial_\beta
H^{[\mu\alpha][\nu\beta]}$ even in flat space, despite the
model's triviality there; no paradox ensues since this pure
superpotential form has vanishing Lorentz generators.] The third
derivatives in the resulting field equations can be removed by
adding the counter-term
\begin{equation}
\Delta I=\int d^5 x\, \varepsilon^{\mu\nu\rho\sigma\tau}
\varepsilon^{\alpha\beta\gamma\delta\epsilon}\,
F_{\mu\nu} F_{\alpha\beta}\,
F^\lambda_{\hphantom{\lambda}\rho} F_{\lambda\gamma}\,
R_{\sigma\tau\delta\epsilon}.
\label{eq18}
\end{equation}
It differs from the action (\ref{eq17}) itself simply by an
overall factor and the index change $\delta \leftrightarrow
\lambda$ in the last two terms. Their sum,
\begin{eqnarray}
I + \Delta I&=&-8 \int d^5 x \sqrt{-g} \Bigl[
4 (F^{\mu\nu} F^{\rho\lambda} F_{\lambda\tau}
F^{\tau\sigma} C_{\mu\nu\rho\sigma})
+ 4 (F^\mu_{\hphantom{\mu}\lambda} F^{\lambda\nu}
F^\rho_{\hphantom{\rho}\tau} F^{\tau\sigma} C_{\mu\rho\nu\sigma})
\nonumber\\
&&+ (F^2) (F^{\mu\nu} F^{\rho\sigma} C_{\mu\nu\rho\sigma})
\Bigr],
\label{eq19}
\end{eqnarray}
depends only on the Weyl tensor $C_{\mu\nu\rho\sigma}$ (for no
obvious $D=5$ reason, though (\ref{eq19}) is manifestly conformal
invariant in $D=10$); as per design, both its $T_{\mu\nu}$ and
field equations depend on at most second derivatives.

\vskip 1.5pc
Details of our models' constructions, of their general
non-minimal compensating gravitational extensions, applications
for instance in the spirit of \cite{EspositoFarese:2009aj}, and
other open questions, e.g., possible supersymmetrization, may be
presented elsewhere.

\section*{Acknowledgments}
The work of S.D. was supported by NSF Grant No.~PHY 07-57190 and
DOE Grant No.~DE-FG02-92ER40701. Our calculations have been
cross-checked using several computer programs, including the
\textit{xTensor} package~\cite{xTensor} for \textit{Mathematica}.

\end{document}